# Clinically Meaningful Explainability for NeuroAI: An ethical, technical, and clinical perspective

Laura Schopp[1], Ambra D'Imperio[1,2], Jalal Etesami[1], Marcello Ienca[1]

[1]Institute for History and Ethics of Medicine, School of Medicine and Health, Technical University of Munich, Germany
[2]Department of Forensic Psychiatry, Universitätsklinik für Forensische Psychiatrie und Psychologie (FPP), Universitäre Psychiatrische Dienste (UPD) Bern, Switzerland

E-mail: laura.schopp@tum.de and marcello.ienca@tum.de

## Abstract

While explainable AI (XAI) is often heralded as a means to enhance transparency and trustworthiness in closed-loop neurotechnology for psychiatric and neurological conditions, its real-world prevalence remains low. Moreover, empirical evidence suggests that the type of explanations provided by current XAI methods often fails to align with clinicians' end-user needs. In this viewpoint, we argue that clinically meaningful explainability (CME) is essential for AI-enabled closed-loop medical neurotechnology and must be addressed from an ethical, technical, and clinical perspective. Instead of exhaustive technical detail, clinicians prioritize clinically relevant, actionable explanations, such as clear representations of input–output relationships and feature importance. Full technical transparency, although theoretically desirable, often proves irrelevant or even overwhelming in practice, as it may lead to informational overload. Therefore, we advocate for CME in the neurotechnology domain: prioritizing actionable clarity over technical completeness and designing interface visualizations that intuitively map AI outputs and key features into clinically meaningful formats. To this end, we introduce a reference architecture called NeuroXplain, which translates CME into actionable technical design recommendations for any future neurostimulation device. Our aim is to inform stakeholders working in neurotechnology and regulatory framework development to ensure that explainability fulfills the right needs for the right stakeholders and ultimately leads to better patient treatment and care.

1. Introduction

Neurotechnology encompasses a wide array of therapeutic and diagnostic tools developed to modulate or interact with the nervous system (1). Its applications can be both implantable or non-implantable and can record brain activity, stimulate it or both at the same time. Cutting-edge neurostimulation leverages adaptive or closed-loop stimulation paradigms that use the available hardware and software to interpret brain signals and to update the stimulation parameters (2). In these paradigms, stimulation parameters are modulated in dependence on internal and external triggers (3).

In recent years, the convergence of neurotechnology and Artificial Intelligence (AI) has paved the way to revolutionize the field and enabling breakthrough innovations (4). From developing personalized Deep Brain Stimulation (DBS) stimulation protocols and real-time stimulation parameter adjustments (4) to facilitating daily seamless speech via a speech neuroprosthesis for patients who are unable to speak (5). Besides DBS, latest AI/Machine Learning (ML) achievements span across all established neurotechnology devices (6). Use cases range from AI copilots supporting BCI users (7), trough AI prediction of neural activity response to responsive neurostimulation (RNS) in epilepsy patients (8), to ML prediction of Vagus nerve stimulation (VNS) response in epilepsy children (9). Neuromorphic devices and hardware exemplify how technology can move beyond conventional ML, demonstrating a high potential for always-on epileptic seizure detection (10).

In closed-loop architectures, the explainability of AI systems reportedly matters to stakeholders (11) including clinicians, patients, developers, and regulators. Technically, there is no unanimous definition of explainability, but there is wide agreement that explainable AI (XAI) models exhibit some ability to shed light on their internal functions, hence generating interpretable of model behavior (12). Such explainability can be generated by applying XAI methods to black-box models or by using inherently interpretable models from the outset. Black box models can be described as models that are (i) too complex for a human to understand, (ii) proprietary, or (iii) both (13). There is a debate on whether medical black box algorithms need to be explainable to be clinically deployed (14–16). We argue that not every AI stakeholder— clinician, patient, developer, and regulator—has the same need and desire for explainability, thus AI explainability needs to be evaluated in a group-specific and setting-specific manner. A large body of literature shows that clinicians, on the one hand, tend to prioritize clinical plausibility (17,18) over explainability (19,20), whereas patients, on the other hand, display heightened explainability needs, as AI opacity is perceived to undermine their autonomy and informed consent (21). Developers (17,22) and regulators (18) likewise exhibit specific explainability needs, reflecting their strong interest in obtaining accounts that allow them to effectively interrogate AI models. Therefore, to bridge the translational gap between AI development, clinical needs, and ultimately therapeutic aims, we provide ethical, technical, and clinical explainability design recommendations to ensure AI-enabled neurotechnology is both socially aligned and clinically meaningful.

## 2. Beyond Explainability: From Technical Transparency to Clinical Meaning

XAI methods aim to make AI more intelligible to humans by generating interpretable insights and explanations into model behavior (23). By increasing transparency, XAI methods are often regarded as a key enabler of trust in AI systems (24,25). Depending on the input data and learning architecture, these explanations may take various forms, including post-hoc assessments like Shapley Additive exPlanations (SHAP). Derived from game theory, SHAP assigns each input feature a numerical value that quantifies its contribution to the model's output (26). For example, SHAP analyses have revealed that preoperative Parkinson's Disease Questionnaire (PDQ-39) scores and upper-beta-band activity are key determinants of quality of life (QoL) outcomes during deep brain stimulation (DBS) for PD (27). Similarly, Local Field Potentials (LFP) serve as important indicators of treatment response in patients with treatment-resistant depression (TRD) (28).

Despite these advances, the real-world adoption of XAI methods in closed-loop clinical neurotechnology remains remarkably low. One recent review found that only about 9% of current studies in AI-enabled neurotechnology for neurology and psychiatry implemented any form of explainability. Among these, six studies employed feature importance techniques, four used SHAP, two unspecified feature-importance analyses, one Gradient-weighted Class Activation Mapping (GradCam), and one unspecified XAI method. This discrepancy between the widespread normative endorsement of explainability as a prerequisite for trustworthy medical AI and its sparse implementation in practice is striking. We argue this may not reflect merely a technical lag but a conceptual misalignment: existing XAI approaches were designed to expose algorithmic structure, not to support clinical reasoning or patient care.

Empirical data show that most clinicians are not interested in exhaustive algorithmic transparency (18). Instead, they value clinically relevant, actionable explanations—for instance, how particular neural features correlate with symptom fluctuations, or how parameter adjustments might influence patient outcomes (18). Full transparency, while theoretically desirable, often overwhelms rather than assists decision-making under time constraints. This divergence underscores the need to move beyond the notion of explainability as mere technical transparency. In clinical neurotechnology, understanding must be relational and context-sensitive. Explanations should be tailored to the epistemic goals, expertise, and workflow of their intended users. A regulator auditing an algorithm's compliance requires a fundamentally different kind of explanation from a neurologist optimizing DBS settings or a patient evaluating treatment options. Hence, explainability in NeuroAI —i.e., AI applied to neurotechnology— must be stratified across ethical, technical, and clinical dimensions. Reorienting XAI toward clinical meaning entails prioritizing actionable interpretability over exhaustive disclosure. Developers should design explanation modules that translate algorithmic inferences into clinically recognizable categories—such as symptom trajectories, stimulation efficacy, or adverse-event probability—while communicating uncertainty transparently. This pragmatic shift aligns with an ethical imperative: to empower clinicians with the right amount of understanding to act safely and responsibly, acknowledging that therapeutic focus and degrees of patient involvement differ across interventions, and that explainability must therefore be tailored to patient-centered goals rather than technical completeness.

Clinically meaningful explainability is therefore not a diluted version of transparency but its purposeful refinement. It seeks to bridge the epistemic distance between machine reasoning and medical judgment, enabling clinicians to integrate AI insights into therapeutic reasoning while retaining epistemic and moral agency. The following section introduces a conceptual framework to operationalize this approach through a tri-dimensional model that integrates clinical, technical, and ethical requirements for explainable NeuroAI.

## 3. Defining Clinically Meaningful Explainability (CME)

If explainability is to be more than a technical ideal, it must be redefined in light of the epistemic and ethical realities of clinical care. In the domain of neurotechnology, where algorithmic outputs can directly shape therapeutic interventions on the brain (29), explainability must serve not only to expose computational processes but to enable safe, responsible, and intelligible action. We propose the concept of clinically meaningful explainability (CME) to describe this reorientation.

Clinically meaningful explainability can be defined as *the capacity of an AI system to provide end-users—particularly clinicians and patients—with explanations that are interpretable, actionable, and relevant within clinical reasoning, thereby enabling justified trust,*

*accountability, and informed decision-making in patient care*. CME is not concerned with rendering every model component transparent but with ensuring that the information conveyed aligns with the clinical, epistemic, and ethical requirements of medical practice. It is thus an instrumental rather than exhaustive form of transparency: its success is measured by how effectively it supports sound clinical judgment and patient well-being.

CME differs from conventional explainability in three key respects. First, purpose alignment: whereas standard XAI seeks to demystify algorithms for technical or regulatory understanding, CME is oriented toward clinical decision support. Its goal is not merely to describe how a model operates, but to clarify why a particular output or recommendation should be trusted and how it can inform therapeutic reasoning. For instance, AI used to recommend the initial DBS parameters for a Parkinson's patient would not merely output the optimal stimulation amplitude, pulse width, and frequency. Instead, a clinically meaningful explanation would specify which patient-specific data informed the recommendation, illustrate how comparable stimulation settings have previously reduced motor symptoms while minimizing side effects, situate these observations within the relevant clinical literature, highlight the patient benefit of the recommendation, and clarify the safety considerations underlying the recommendation. Second, cognitive fit: CME emphasizes interpretive formats that correspond to the clinician's workflow, knowledge base, and perceptual intuitions. For instance, a neurologist may benefit from a graphical representation linking neural feature trends to symptom trajectories rather than a complex map of internal network weights. Third, ethical adequacy: CME explicitly integrates values such as autonomy, accountability, and nonmaleficence. An explanation that is technically precise but undermines clinician confidence or patient understanding fails to meet the ethical standard of meaningfulness. Additionally, CME provides the basis for a corrective mechanism when the AI recommendation diverges from the clinician's judgment. As it reframes explainability as a relational and normative property, CME embeds the explanatory process within the ecology of medical decision-making. Explanations in NeuroAI are not delivered into a vacuum but unfold within clinical dialogues, institutional routines, and moral responsibilities. CME, therefore, cannot be engineered at the algorithmic level alone; it must emerge from co-design between clinicians, engineers, ethicists, and patients. Through such collaboration, CME bridges the epistemic logic of machine learning with the practical logic of medicine and calls for calibrating the depth and format of explanations to user needs. A regulator may require traceable model documentation, a clinician a visualized confidence interval for predicted outcomes, and a patient a concise rationale to support informed consent. Tailoring the form of explainability to its function transforms transparency from an abstract ideal into a pragmatic design ethos. Understanding, in this view, does not entail reconstructing every computational step but attaining sufficient epistemic and moral clarity to act responsibly. The next section introduces a tri-dimensional model that operationalizes CME across clinical, technical, and ethical domains.

## 4. A Tri-Dimensional Model for Clinically Meaningful Explainability in NeuroAI

Translating CME from principle into practice requires a framework that connects ethical intent, technical feasibility, and clinical utility. To this end, we propose a tri-dimensional model for CME in NeuroAI. This model conceptualizes explainability as emerging from the dynamic interaction among three interdependent domains—clinical, technical, and ethical—each defining a distinct yet overlapping set of goals, constraints, and responsibilities.

*The Clinical Dimension (Actionable Interpretability):* The clinical dimension grounds explainability in the realities of diagnosis, treatment, patient monitoring, and focusing more on the patient-centred therapeutic approach. Clinicians operate under cognitive, temporal, and informational constraints that make exhaustive transparency impractical. Instead, they require explanations that are actionable, context-aware, and integrated into clinical reasoning. This involves representations that link algorithmic inferences to medically meaningful categories—such as symptom trajectories, biomarkers of therapeutic response, or confidence levels associated with stimulation adjustments. For example, in adaptive Deep Brain Stimulation (aDBS), the system's modulation of parameters based on neural activity should be explainable in a format that clarifies how changes in oscillatory power or network synchrony relate to motor improvement or side effects, and how these relate to the best patient-specific therapeutic target, while respecting both the subjective and ethical dimensions of clinical decision-making. Such grounded interpretability allows the clinician to justify therapeutic decisions to colleagues, regulators, and, more importantly, patients while maintaining epistemic agency. In this domain, explainability is successful not when it reveals the inner workings of the algorithm, but when it facilitates accurate, timely, and ethically defensible medical practice.

*The Technical Dimension (Computational Fidelity):* The technical dimension ensures that clinically meaningful explanations are scientifically valid and computationally robust. Generating interpretable outputs must not compromise the predictive accuracy, stability, or latency of the underlying model—especially in real-time, safety-critical systems such as closed-loop neurostimulators. Here, explainability must be implemented through *fidelity-preserving*

*mechanisms*: for instance, using post-hoc feature-attribution models that approximate neural signal–output relationships without interfering with system responsiveness, or employing hybrid architectures that combine interpretable surrogate models with high-performance black boxes. The technical domain also encompasses design strategies for *adaptive explainability*: systems that modulate the depth or modality of their explanations depending on the user role, task urgency, or available data. Advances in neuromorphic computing may further support these goals by enabling low-latency low-power computing for the next generation of neurotechnology (30). In this sense, the technical dimension of CME is not ancillary to performance—it is an integrated design principle that balances intelligibility and efficacy.

*The Ethical Dimension (Value Alignment and Accountability):* The ethical dimension anchors CME in the normative commitments of medicine: respect for autonomy, beneficence, and accountability. An explanation is ethically meaningful when it enables responsible action—when it allows a clinician to justify decisions to patients, regulators, and society. This dimension requires that explanations be accessible to all relevant stakeholders, including patients, without generating undue cognitive burden or false confidence. It also demands that responsibility remain traceable: clinicians should understand the extent of algorithmic agency, and patients should know whether decisions arise from human judgment, machine recommendation, or a hybrid process. From a governance perspective, embedding ethical values into XAI design implies establishing standards for explanatory sufficiency—the minimal level of interpretability required for informed consent, risk communication, and post-hoc accountability. In this regard, CME aligns with recent policy developments, such as the EU AI Act's provisions on "appropriate transparency" for high-risk systems, but specifies how this transparency should be operationalized in clinical contexts involving the brain.

## 5. Proposed CME reference system architecture

To illustrate how CME can be implemented, we outline a reference architecture called NeuroXplain. To enhance actionability, we integrate NeuroXplain into Medtronic's RC + S device system architecture, which enables researchers to record and stimulate neural activity and to configure a stimulation algorithm for research and investigational purposes, as well as for human subjects. The concrete elements are inspired by expert interview findings (18) and based on expert knowledge. Medtronic's RC + S device system architecture consists of several elements: Data Management, Host Programmer, Research Programmer, Telemetry Transceiver, Recharger, Patient Programmer, Neurostimulator, Extension, Bifurcated Extension, 1x4 lead (3x).

It uses an 8-dimensional linear-SVM detector (top) (see 31). For the operationalization of CME, we consider the Summit RC+S device, which is used to treat Parkinson's patients and would add a new component to the existing system: an XAI dashboard (left). In the following, we operationalize CME for each system element by discussing it along the clinical, technical, and ethical dimensions.

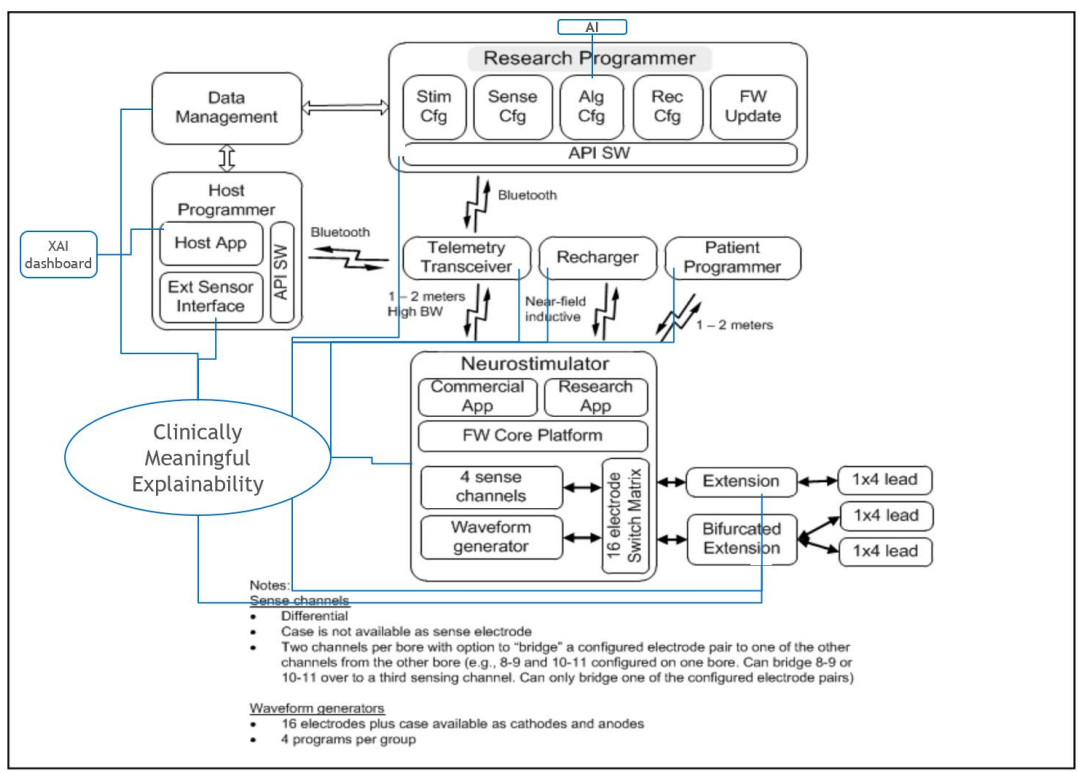

Figure 1: RC+S System architecture including CME elements. The Summit RC+S system allows investigational research in the field of neuromodulation for human subjects (31). Adapted from (32).

*5.1. The Clinical Dimension:*

**XAI dashboard**

*CLINICAL:* Clinicians must interpret AI-generated recommendations. To do so responsibly, they require information regarding both the input data used to generate a prediction and the rationale underlying the output. Our proposed AI dashboard would allow clinicians to find all clinically relevant information about the whole AI pipeline. Specifically, clinicians would benefit from access to the respective LFP signals for a specified period, their corresponding timestamps, patient activity data, and the current stimulation parameters. Whenever available, they would also receive input data from connected neurotechnology devices, e.g., symptom detection data for patients with PD. Clinicians will then be able to clearly see on the dashboard the input parameters, such as beta power and symptoms like motor rigidity. Concerning the output, clinicians will acquire insights into whether the AI's recommendation is supported by existing clinical evidence, the potential side effects of the suggested simulation, and its expected efficacy. This will occur in full compliance with the EU AI Act. See **Figure 2**.

*TECHNICAL:* The dashboard should enable the technician to select from different models and, for each selected model, display all relevant outputs in the requested format (e.g., curves, tables), along with the corresponding model parameters. This functionality allows the technician to experiment with different models and parameter configurations in order to identify the best-performing solution. Another important feature of most machine-learning-based methods is access to user feedback. Therefore, it is important that the dashboard enables clinicians to provide feedback in a concise, structured manner. This feedback can subsequently be incorporated into the learning algorithm as expert-provided labels, thereby improving model performance over time.

*ETHICAL:* The dashboard helps to promote transparency, accountability, and ultimately trust. End-users must clearly see all available information, including the proven benefits of the AI recommendation. By making all necessary elements visible, AI-enabled decisions remain understandable to end users and thus comply with regulatory requirements for high-risk systems such as DBS under the EU AI Act.

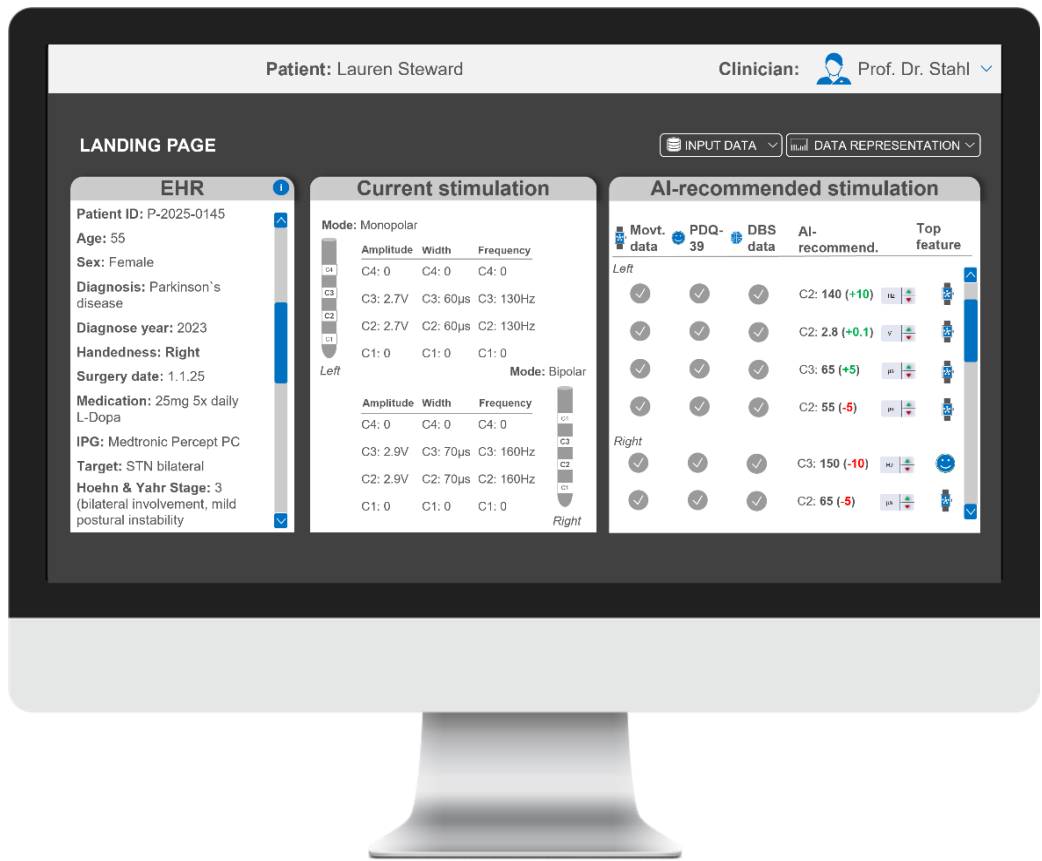

**Figure 2:** Demonstration of XAI dashboard with dummy data and inspired by (18).

**Data Management**

*CLINICAL:* Neural data (e.g., LFPs, electroencephalogram (EEG), or electrocorticography) is ideally recorded 24/7. Data storage must comply with the General Data Protection Regulation and avoid cloud services when possible. Integration with electronic health records to ensure medically meaningful datasets across different disciplines such as neurology, psychiatry, and neurosurgery.

*TECHNICAL:* The most important component of any AI-based system is its training dataset. Therefore, from a technical perspective, data recording and management are crucial for our CME framework. Since medical data is safety-critical, the recording procedure must be as accurate as possible and include an audit mechanism to ensure the validity of the data, particularly when recordings are disrupted by patient actions or sensor failures. All data should be encrypted during recording, transmission, and storage to preserve privacy. Especially the data flow between the neurostimulator and the telemetry transceiver, recharger, and the patient programmer. Access to the data should be strictly controlled based on the roles and responsibilities of clinicians and technicians. When recorded data is not clean, pre-processing often accounts for more than 60% of the data analysis process. Consequently, our data recording and storage procedures should produce data that is as clean as possible, especially when near real-time outputs are required. To achieve this, various subroutines should be integrated into the data collection pipeline to monitor data quality. For example, in cases of inaccurate measurements, due to sensor failure or signal power falling below a threshold, data points should be automatically excluded and logged. These subroutines can also perform preliminary data analysis and compression to save memory and reduce computational complexity.

Once the data has been properly collected and pre-processed, technicians can apply advanced models for analysis and knowledge extraction, including a post-hoc feature-attribution model and causal reasoning approaches. Recent advances in machine learning and causal inference allow us to distinguish correlation from causation, partially recovering the underlying causal relationships among system variables (33). Such causal knowledge facilitates human-understandable explanations, for example, "Diagnosis X is indicated because Y occurs and Z follows" and helps identify root causes of anomalies within the data (34). Applying multiple models, such as Multivariate Hawkes processes (35,36), Hidden Markov Models (37), Generalized Linear Models (38), or deep neural networks, can improve overall accuracy by allowing clinicians to use combination strategies, such as majority voting or following the best-expert strategy (39). Furthermore, since each model can capture different aspects of the data, their combined use provides technicians with a more comprehensive explanation of the observed patterns. Finally, the results of these AI-based analyses can be presented in a user-friendly format using large language models (LLMs). Finally, a feedback loop should be implemented. As noted earlier, access to reliable labelled data is crucial for improving AI performance. Therefore, after the AI provides its recommendation, clinicians' feedback (including matches and mismatches) should be collected and stored along with the relevant model parameters to support continual learning and system refinement.

*ETHICAL:* All collected data constitute neural data, a category now explicitly recognized and adopted by all UNESCO member states in their final Draft Recommendation on the Ethics of Neurotechnology (29). As such, they require the highest level of protection to safeguard mental privacy. Data management must ensure compliance with all national regulations. Raw neural data must be appropriately anonymized to prevent individual re-identification. Patients should consent to the storage and processing of the data.

**Patient Programmer**

*CLINICAL:* Patients must be able to switch off the device in case of discomfort to avoid any harm and promote autonomy. The interface will therefore communicate the therapy status in plain language that does not require any AI literacy.

*TECHNICAL:* Patients should be able to turn sensors on or off, or reduce the intensity of the measuring devices (where applicable) whenever they experience discomfort. However, any changes to the sensors' settings should be recorded, as they indicate periods during which the data may be missing or altered in intensity. Technically, this means that knowing these intervals can help technicians more accurately analyse the recorded time series.

*ETHICAL:* Patient enablement is operationalized through patient-centred features that foster trust, autonomy, and transparency. Patients should have the final possibility to switch the device off. However, depending on their condition, they should not be confronted with decisions about the stimulation paradigm.

**Research Programmer**

*CLINICAL:* Clinicians will remain in the decision loop, especially at the early stages of AI development. AI recommendations can appear as recommendations that require clinician approval at routine stimulation setting meetings. A feedback module for the AI recommendations will allow clinicians to label them as "clinically plausible", unclear", or "irrelevant", which in turn allows the system to improve prediction in the long run.

*TECHNICAL:* To ensure safety in AI-assisted decision-making, clinicians should remain actively in the decision loop. Technically, this means that the differences between the AI's recommendations and the clinician's decisions should be recorded. To encourage clinicians to provide feedback, instances where both recommendations coincide should be automatically labelled as "matched." When recommendations differ, a predefined set of response options should be presented for the clinician to select from, with the option to provide additional explanations. Another important feature that can help technicians improve the AI's performance is to record all mismatched scenarios, including the corresponding input data and all relevant model parameters. Such scenarios can be used for retraining and ensuring reproducibility.

*ETHICAL:* Best practice requires continuous documentation and monitoring of AI accuracy and safety. Confidence intervals should reflect model uncertainty and the reliability of recommendations. Audit trails should document model recommendations for transparency purposes and clinician overrides to preserve accountability. Oversight by a human supervisor must remain integral to decision-making and any adjustments to stimulation parameters.

**Recharger**

*CLINICAL:* Recharger systems should be optimized for long battery life and stability to reduce the need for surgical replacements and minimize infection risks. Battery design significantly influences the IPG size. Balancing energy efficiency and miniaturization is therefore critical to further improve patient experience and comfort.

*TECHNICAL:* The recording procedure should be precise and technically robust. To achieve this, measures such as using sensors with rechargeable batteries, employing multiple sensors when feasible, and utilizing sensors with diagnostic capabilities, such as wearable devices for daily monitoring, can be implemented.

**Extensions**

*CLINICAL:* Clinicians should be able to choose between standard and bifurcated extensions depending on the treatment plan and the patients' needs. All leads must be CE-certified for DBS in Parkinson's disease. In addition, clinicians require clear documentation of the safety profile of the extensions, including MRI compatibility and any associated procedural limitations.

## 6. Operationalizing Clinically Meaningful Explainability in Practice in NeuroAI systems

To translate clinically meaningful explainability (CME) from concept to implementation, NeuroAI systems must be designed, evaluated, and governed in ways that reflect the needs of their primary users—clinicians and patients—while preserving scientific integrity and ethical accountability. Operationalizing CME requires coordinated action across three complementary domains: design, validation, and governance.

*Design:* Building Explainability into the Clinical Workflow: Explainability should not be treated as a post-hoc feature added after model training, but as a co-constitutive element of system design. Developers should adopt human-in-the-loop and value-sensitive design approaches that involve clinicians and patients early in the development process to identify what kinds of explanations are most meaningful in context. For example, iterative co-design workshops can help translate complex model outputs—such as neural signal variance, confidence scores, or feature importance—into clinically intuitive visualizations. Interactive dashboards that map AI predictions to symptom trajectories or stimulation adjustments can support decision-making without overwhelming users with unnecessary technical details.

Furthermore, design efforts should prioritize adaptive explainability: systems that modulate the depth and form of explanations depending on user profile and context. A clinician performing real-time parameter tuning may need rapid visual summaries of algorithmic reasoning, whereas a regulator might require a full audit trail of model performance and training data. Embedding CME in this adaptive manner ensures that explainability remains both relevant and scalable across different use environments.

*Validation:* Operational explainability must also be empirically validated. Current AI evaluation pipelines focus on accuracy, sensitivity, and specificity, but rarely assess whether explanations improve understanding, decision quality, or patient outcomes. CME calls for the integration of explainability metrics into clinical validation studies. These should measure not only the technical fidelity of explanations (e.g., how faithfully they represent model logic) but also their cognitive and practical utility for end-users. Quantitative measures—such as task performance under time constraints, decision confidence, or diagnostic consistency—can complement qualitative assessments derived from structured interviews or think-aloud protocols. Importantly, CME validation must be domain-specific: the same form of explanation that aids understanding in DBS may confuse users in neuroprosthetics or brain–computer interface contexts. Developing standardized yet flexible assessment frameworks will therefore be crucial for benchmarking meaningful explainability across applications.

*Governance:* The operationalization of CME extends beyond design and testing—it must be embedded in institutional and regulatory frameworks that ensure accountability. Regulators should move beyond generic calls for transparency and establish explainability sufficiency criteria tailored to clinical contexts. These criteria should specify the minimal level of interpretability required to justify trust, support informed consent, and enable post-hoc accountability in the event of malfunction or harm. Ethics committees and hospital review boards can play an important role by requiring CME documentation in the evaluation of NeuroAI systems. This may include clear statements of who the explanations are for, how they are generated, and what epistemic limitations they carry. Moreover, continuous monitoring mechanisms should be established to assess whether explanations remain meaningful as systems evolve through adaptive learning or software updates. Such dynamic oversight reflects the principle that explainability, like clinical safety, is not a one-time achievement but an ongoing responsibility.

## 7. Conclusion

Explainability has long been regarded as a cornerstone of trustworthy AI, yet in clinical neurotechnology, its promise has remained largely theoretical. As AI-enabled systems increasingly participate in decisions that modulate brain function, the need for transparency is not merely epistemic but moral. The capacity to understand why an algorithm recommends a given intervention is integral to clinical responsibility, patient autonomy, and public trust. However, the pursuit of maximal transparency has often obscured a more urgent question: what kind of understanding do clinicians and patients need to act wisely? Clinically meaningful explainability offers a pragmatic and ethically grounded answer. By reorienting explainability from model introspection toward clinical reasoning, CME recognizes that understanding in medicine is context-dependent and action-oriented. The tri-dimensional model we propose—integrating clinical, technical, and ethical domains—shows that meaningful explainability arises not from any single feature but from the balance between interpretability, fidelity, and accountability. In practice, this means designing AI systems that deliver explanations fitted to the user's cognitive and moral task, validating them for actual clinical benefit, and embedding them within governance frameworks that uphold transparency and trustworthiness as evolving responsibilities. Our proposed system, NeuroXplain, can redefine success in NeuroAI from building systems that can be explained in principle to designing systems that can be understood and trusted in use. By operationalizing CME across design, validation, and governance, we move closer to a future where AI-enabled neurotechnologies are not only powerful and adaptive but also epistemically, ethically, and clinically aligned.

Policy initiatives such as the EU AI Act, the U.S. FDA's Good Machine Learning Practice guidelines, and WHO's AI ethics principles provide a foundation for this shift but remain largely silent on the question of meaningfulness. Integrating CME into these frameworks would help ensure that regulatory demands for transparency translate into real clinical utility rather than bureaucratic formality. Likewise, medical institutions and ethics committees should treat explainability not as a technical add-on but as an ethical criterion of design and deployment. Ultimately, the success of NeuroAI will not depend solely on computational performance but on the intelligibility of its decisions to those who bear their consequences. Clinically meaningful explainability reframes transparency as a clinical virtue—one that enables informed action, preserves professional judgment, and aligns the logic of machines with the values of medicine. In doing so, it lays the foundation for an ethically robust and human-centered future of neurotechnology.

## Declarations

**Consent for publication**

All authors contributed to this article and approved the submitted version.

**Competing interests**

LS is an employee at Boston Consulting Group GmbH. MI has been a policy advisor on neurotechnology to the OECD, the Council of Europe, the UN, and the EU Parliament. He is a member of the ethics board of IDUN Technologies, a company producing EEG-earbuds.

**Funding**

This study has been funded by Hightech Agenda Bayern (PI: MI). LS's PhD scholarship has been funded by Boston Consulting Group GmbH.

**Acknowledgment**

We thank all our interviewees for their time and willingness to participate in our study.

We acknowledge using Grammarly and ChatGPT (GPT-4o) for proof-reading purposes.

**ORCID Ids**

L Schopp https://orcid.org/0009-0001-1168-4896

A D'Imperio https://orcid.org/0000-0002-7692-8986

J Etesami https://orcid.org/0000-0002-7404-5952

M Ienca https://orcid.org/0000-0001-8835-5444